
\input phyzzx.tex

\def\b{\beta}

\def\d{\delta}

\def\e{\epsilon}

\def\o{\omega}
\def\O{\Omega}

\def\pa{\partial}

\def\bbox{{\,\lower0.9pt\vbox{\hrule \hbox{\vrule height 0.2 cm
\hskip 0.2 cm
\vrule height 0.2 cm}\hrule}\,}}
%

\def\pl#1{{\it Phys. Lett.} {\bf B#1}}
\def\prl#1{{\it Phys. Rev. Lett.} {\bf #1}}
\def\prd#1{{\it Phys. Rev.} {\bf D#1}}

\def\np#1{{\it Nucl. Phys.} {\bf B#1}}

\def\mpl#1{{\it Mod. Phys. Lett.} {\bf A#1}}

%
\REF\thooft{G. 't Hooft, \np {256}, 727 (1985).}
\REF\call{C. Callan and F. Wilczek, \pl {333}, 55 (1994);
D. Kabat and M. J. Strassler, \pl {329}, 46 (1994).}
\REF\suss{L. Susskind and J. Uglum, \prd {50}, 2700 (1994).}
\REF\barb{J. L. F. Barbon, \prd {50}, 2712 (1994).}
\REF\emp{R. Emparan, preprint EHU-FT-94/5, hep-th/9407064.}
\REF\frolov{V. Frolov and I. Novikov, \prd{48}, 4545 (1993).}
\REF\solod{S. N. Solodukhin, preprints hep-th/9407001,
hep-th/9408068.}
\REF\ghm{A. Ghosh and P. Mitra, \prl{73}, 2521 (1994).}
\REF\bek{J. Bekenstein, \prd {7}, 2333 (1973); {\bf D9}, 3292
(1974).}
\REF\hawk{G. Gibbons and S. Hawking, \prd{15}, 2752 (1977).}
\REF\teit{C. Ba\~{n}ados,C. Teitelboim, and J. Zanelli, \prl{72}, 957
(1994); C. Teitelboim, preprint hep-th/9410103.}
\REF\gp{G. Gibbons and M. Perry, {\it Proc. R. Soc. Lond.} {\bf
A358},
467 (1978).}
\REF\bd{N. D. Birrell and P. C. W. Davies, Quantum fields in curved
space, Cambridge Univ. Press, 1982.}
\REF\gid{S. Giddings, \prd{49}, 4078 (1993) }
\REF\hor{G.W. Gibbons, in Fields and Geometry, proceedings of 22nd
Karpacz
Winter School   1986, ed. A. Jadczyk (World Scientific, 1986), H. F.
Dowker,
J. Gauntlett,  S. B. Giddings,and G. Horowitz, \prd{50} 2662 (1994),
S.
Hawking, G. Horowitz and S. Ross, preprint NI-94-012,
gr-qc/9409013, and C. Teitelboim in reference 11. }
\REF\gm{G. Gibbons and K. Maeda, \np{298}, 741 (1988); D. Garfinkle,
G. Horowitz, and A. Strominger, \prd{43}, 3140; {\bf D45}, 3888 (E)
(1992).}
\REF\preskill{J. Preskill, P. Schwarz, A. Shapere, S. Trivedi, and
F. Wilczek, \mpl{6}, 2353 (1991).}
\REF\bom{L. Bombelli, R. Koul, J. Lee and R. Sorkin,
\prd {34}, 373 (1986).}
\REF\sred{M. Sredinicki, \prl {71}, 666 (1993).}

\pubnum {COLO-HEP-347, OU-HET 207 \cr hep-th/9412027}
\date={November 1994}
\titlepage
\vglue .2in
\centerline{\bf On the Entropy of Quantum Fields in Black Hole
Backgrounds}
\author{ S.P. de Alwis\foot{e-mail: dealwis@gopika.colorado.edu} }
\address{Dept. of Physics, Box 390,\break
University of Colorado,\break Boulder, CO 80309}
\author{N. Ohta\foot{e-mail: ohta@phys.wani.osaka-u.ac.jp}}
\address{Dept. of Physics, Osaka University,\break
Toyonaka, Osaka 560, Japan}
\vglue .2in
\centerline{\caps ABSTRACT}
We show that the partition function for a scalar field in a static
spacetime background can be expressed as a functional integral
in the corresponding optical space, and point out that the difference
between this and the functional integral in the original metric is
a Liouville type action. A general formula for the free
energy is derived in the high temperature approximation and applied
to various cases. In particular we find that thermodynamics in the
extremal Reissner-Nordstr\"om space has extra singularities that
make it ill-defined.

\endpage

Recently much  attention has been paid to the calculation of the
quantum corrections [\thooft-\ghm] to the Bekenstein-Hawking entropy
[\bek,\hawk] of black holes. In this paper, we will derive a general
formula for the free energy and entropy of a scalar field in an
arbitrary static spacetime background in the high temperature
approximation. We will show that the difference between our free
energy and that in the calculations of [\call,\suss] is due to
the conformal anomaly. We will also apply our formula to various
cases
and in particular discuss a possible
resolution of a puzzle associated with the thermodynamics of extremal
Reissner-Nordstr\"om black holes.

Consider a static metric $ ds^2= g_{00}dt^2 + h_{ij}dx^idx^j$.
Writing
$g=\det g_{\mu\nu}=g_{00}h,~h=\det h_{ij}$ where $\mu ,\nu
=0,...,D;~i,j=1,...D$, we have the action for scalar fields in this
background
$$
\eqalign{S
&=-{1\over 2}\int d^Dx\sqrt{-g}g^{\mu\nu}\pa_{\mu}\phi\pa_{\nu}\phi
\cr
&=\int dt\int d^{D-1}x\sqrt h [{1\over
2\sqrt{-g_{00}}}\dot\phi^2-{\sqrt{-g_{00}}\over
2}h^{ij}\pa_i\phi\pa_j\phi ].}
\eqn\action
$$
The canonical momentum is $\pi={\dot\phi\over \sqrt{-g_{00}}}$ and
the Hamiltonian is
$$
H=\int d^{D-1}x{\cal H}=\int d^{D-1}x\sqrt h \sqrt{-g_{00}}[{1\over
2}\pi^2+{1\over 2} h^{ij}\pa_i\phi\pa_j\phi ],
\eqn\ham
$$
and the equal-time canonical commutation relations are
$[\hat\phi (\vec{x}),\hat\pi (\vec{y})]={i\over\sqrt h}\d
(\vec{x}-\vec{y}).$
By the usual time-slicing method, one finds for the partition
function
in this background the expression
$$
\eqalign{\Tr [e^{-\b H}]&=\int [d\pi]\int_{\phi (0,\vec{x})=\phi
(\b,\vec{x})} [d\phi]e^{-\int_0^{\b}dt \int d^{D-1}x\sqrt{h}
[-i\pi\dot\phi+{\cal H}]}\cr &=\int_{\phi
(0,\vec{x})=\phi(\b,\vec{x})}
\prod_{t,\vec{x}}^{}{d\phi\left({h\over g_{00}^E}(t,\vec{x})
\right)^{1\over 4}}e^{-\int_0^{\b}dt \int  d^{D-1}x\sqrt{g^E}
{1\over 2}g^{E,\mu\nu}\pa_{\mu}\phi\pa_{\nu}\phi} .}
\eqn\part
$$
In the above $g_{\mu\nu}^E=(-g_{00}, h_{ij})$ is the Euclidean metric
and henceforth we will drop the superscript $E$. It is convenient to
discuss conformally coupled scalars and to introduce a
mass term, so we will change the matter action
(after partial integration) to $S_{\phi}=\int _0^{\b}dt
d^{D-1}x\sqrt{g}\phi(K+m^2)\phi,$ where
$K\equiv -\bbox + {1\over 4}{D-2\over D-1}R,~~\bbox \equiv
{1\over\sqrt{g}}\pa_{\mu}(\sqrt{g}g^{\mu\nu}\pa_{\nu})$.
Thus we may write
$$
\eqalign{\Tr [e^{-\b H}]&=\int_{\phi (0,\vec{x})=\phi (\b ,\vec{x})}
\prod_{t,\vec{x}}^{}{d\phi \O g^{1\over 4} (t,\vec{x})}
e^{-\int_0^{\b}dt d^{D-1}x\sqrt{g}\phi (K+m^2)\phi}\cr
&=\int_{\phi (0,\vec{x})=\phi(\b,\vec{x})} \prod_{t,\vec{x}}^{}{d\phi
g^{1\over 4}(t,\vec{x})}e^{-\int_0^{\b}dt d^{D-1}x\sqrt{g}
\phi (K+m^2)\phi +S_L[g, \O]} .}
\eqn\partition
$$
In the above $\O ={1\over\sqrt {g_{00}}}$ is a conformal factor which
causes a mismatch between the metric background of the action and
that
defining the functional integral.  The effect of
this term may be written as a Liouville type action. In two
dimensions, it is in fact the Liouville action with the Liouville
field being $\ln\O^2$.

Thus we have for the free energy the expression
$$
-\b F=-{1\over 2}\ln\det [K_{\b}+m^2]+\b\int d^{D-1}x \sqrt g
L_{L}[\O, g].
\eqn\free
$$
The second term is linear in $\b$ so that the temperature dependence
of the free energy and hence the entropy ($S= \b^2{\pa F\over\pa\b}$)
comes entirely from the first term.
Away from the Hawking temperature the Euclidean metric has
conical singularities with $\int R\sim \b_{Hawking}-\b $
[\teit,\suss,\call]. However these $\beta$-dependent terms vanish
at the Hawking temperature and the bulk term is simply
the quantum correction to the zero temperature cosmological constant
\foot{This would be zero in a supersymmetric theory.}
which should be canceled against a bare cosmological constant.
Hence the entire free energy of the gas of particles at the Hawking
temperature must come from the generalized Liouville action.

There is, however, a formulation of the path integral in which the
calculation is directly related to the evaluation of the free energy
of a gas of bosons. This is obtained by introducing the
optical metric [\gp]\foot{This metric has been used in connection
with this problem also in [\barb,\emp]. But unlike in those papers
here we show how this metric arises from the standard expression
for the partition function.}
and performing a change of field variable. Thus writing
$\bar g_{\mu\nu}=\O^2 g_{\mu\nu},~~\bar\phi = \O^{2-D\over 2}\phi$,
we have for the measure $ \prod_{t,\vec{x}}^{}{d\phi \O g^{1\over 4}
(t,\vec{x})}=\prod_{t,\vec{x}}^{}{d\bar\phi  \bar g^{1\over 4}
(t,\vec{x})}$. Using the properties of the Laplacian with conformal
coupling under a conformal transformation (see for example [\bd]),
we may write the partition function as
$$
\eqalign{\Tr [e^{-\b H}]=&\int_{\bar\phi (0,\vec{x})=\bar\phi
(\b,\vec{x})}\prod_{t,\vec{x}}^{}{d\bar\phi  \bar g^{1\over 4}
(t,\vec{x})}e^{-\int_0^{\b}dt d^{D-1}x\sqrt{\bar g}\bar\phi (\bar K
+m^2\O^{-2})\bar\phi}\cr =&=-{1\over 2}\ln\det [\bar K_{\b}
+m^2\O^{-2}]=-\int_{\e}^{\infty} {ds\over s}\int\sqrt {\bar g} d^Dx
\bar H(s|x,x).}
\eqn\partopt
$$
Here $\bar H(s|x,x') = e^{-s (\bar K+m^2\O^{-2})}{1\over \sqrt{\bar
g}}
\d^{D}(x-x')$ is the heat kernel and $\e$ is an ultraviolet cutoff.
Optical space has the metric $\bar ds^2=dt^2+{h_{ij}\over g_{00}}dx^i
dx^j$, and it has the topology $S^1\times {\cal M}^{D-1}$, so that
the heat kernel factorizes into that on $S^1$ and the one on
${\cal M}^{D-1}$. Hence we have the following formula for the
free energy after subtracting the zero-temperature cosmological
constant term (i.e. the $n=0$ term in the thermal sum):
$$
F(\b )=-{1\over 2}\int_0^{\infty}{ds\over s}
{1\over (4\pi s)^{D\over 2}}\sum_{n\ne 0}
e^{-{\b^2n^2\over 4s}}\sum_{k=0}^{\infty}{(-s)^k\over k!}\bar B_k.
\eqn\freeenergy
$$
The first factor in the integral is the heat kernel on $S^1$ and in
the second factor we have used the well-known expansion for the heat
kernel [\bd] with
$$
\bar B_0=\int_{\cal M}e^{-\O^{-2}m^2s}\sqrt {\bar g}, ~~\bar
B_1=(\xi-{1\over 6})\int_{\cal M}e^{-{\O^{-2}m^2}s}\sqrt{\bar g}{\bar
R},
\eqn\bzero
$$
etc., where $\xi ={1\over 4}{D-2\over D-1}$.
It should be noted that the free energy has the expected ultraviolet
divergence, but it does not come from the $s=0$ end of the proper
time integral. Instead it is the divergence of the optical
metric at a horizon of the original space that causes trouble. We
will discuss this further by looking at particular examples,
but before that let us derive a universal expression for
all static spaces by using the high temperature approximation. This
is easily obtained by first changing the variable of the proper
time integral from $s$ to $u=\b^{-2}s$ and then neglecting the higher
powers of $\b^2$ coming from the expansion in \freeenergy:
$$
\eqalign{
F=&-T^DV_{D-1}\int_0^{\infty}{du\over u}
{1\over (4\pi u)^{D\over 2}}\sum_{n= 1}^{\infty}e^{-{n^2\over 4u}}
\cr
=& -{T^D V_{D-1} \over {\pi^{D \over 2}}} \Gamma\left({D\over
2}\right) \zeta(D),}
\eqn\highfree
$$
where $V_{D-1}=\int_{ { \cal M_{\rm D-1}}}\sqrt{\bar g}$ is the
volume of optical space. This is just the free energy of a gas of
(massless) particles in a box whose volume is given by the optical
measure. Thus in four dimensions we have
$$
F=-V_3 T^4{\pi^2\over 90}.
\eqn\fourdfree
$$
Note that these formulae for the free energy are physically
relevant only at the Hawking temperature but we need these
expressions at arbitrary $T$ to calculate the entropy of the quantum
fields from the relation $S=-{\pa F\over\pa T}|_{T=T_H}$.

Let us now discuss some examples. The first is 2D Rindler space.
The entropy has been calculated by several authors [\call,\suss]
using the path integral in the original metric \partition, but the
Liouville action term of this equation was not kept, so that the
free energy at the
Hawking temperature was not obtained by them.
Let us check that this term indeed gives the right value for the
free energy of massless particles. Euclidean Rindler space has the
metric $ds^2
=R^2 d\o^2
+dR^2 $ and the relevant conformal factor in \partition\ is $\O
={1\over R}$. The Hawking temperature $T_H$ is ${1\over 2\pi}$, so
at this value the free energy is given by the Liouville action.
Thus we have\foot{For a related calculation see [\gid].}
$$
-2\pi F= S_L={1\over 24\pi}\int d^2x\sqrt{g}g^{\mu\nu}\pa_{\mu}
\ln\O\pa_{\nu}
\ln\O={1\over 24\pi}\int_0^{2\pi}\int_{\e}^{L} RdR {1\over R^2},
\eqn\1
$$
and the free energy is $F=-{1\over 24\pi}\ln{L\over\e}$. This agrees
with what one gets by using the optical metric formulation in which
case one has the exact result \highfree\ (since all
curvature terms are zero) that
the free energy is that of a gas of (massless) bosons in a box of
optical
volume
$V_1=\int_{\e}^L {dR\over R}$ at $T_H={1\over 2\pi}$. In dimensions
greater than two, however, the optical curvature is non-zero
($\bar {\cal R}=-(D-1)(D-2)$) and one has to use the high temperature
approximation, i.e. \highfree\ with
$V_{D-1}=V_{D-2}\int_{\e}^{\infty} {dR\over R^{D-1}}={V_{D-2}\over
(D-2)\e^{D-2}} $. Thus in four dimensions we find (using
\fourdfree) $F=-{A\over \e^2} T^4{\pi^2\over 180}$, where $A$ is
the transverse area in agreement with the first calculation of
[\suss].

We will now discuss (four-dimensional) black hole spaces. The
Schwarzschild metric (setting $G_N =1$) is $ds^2 =- (1-{2M\over r})
dt^2 + {dr^2\over (1-{2M\over r})}+r^2 d\O_2$ and the corresponding
optical volume is
$$
\eqalign{V_3^{Sch}&=4\pi\int^R_{2M+\e}{r^2\over (1-{2M\over r})^2}dr
\cr
&=4\pi [{R^3\over 3 }+2MR^2 +12M^2R+32M^3\ln{R-2M\over\e}\cr
&+{16M^4\over\e}-{104M^3\over 3}+O(R^{-1})+O(\e )].}
\eqn\schvol
$$

By plugging this into \highfree, we immediately get the free energy
and hence
the entropy
of a scalar field in a black hole background. Here we see
the divergence first observed by [\thooft]. Although it appears
linear
in terms of the coordinate cutoff $\e$, it is quadratic in terms of
the proper distance cutoff $ \d=\sqrt{2M\e}$ in the Schwarzschild
geometry. We also see another logarithmic divergence.
These additional divergences can also be  found [\solod] by working
with the functional integral in the original metric \partition.
However in that case the calculation is much more complicated.

Next let us consider the Reissner-Nordstr\"om charged black hole.
This example is interesting because it has an extremal limit when
the mass becomes equal to the charge. The metric is
$$
ds^2=-(1-{2M\over r}+{Q^2\over r^2})dt^2+(1-{2M\over r}+{Q^2\over
r^2})^{-1}dr^2+r^2d\O_2.
\eqn\rn
$$
This black hole has an ADM mass $M$ and an electric charge $Q$.
The metric has outer and inner horizons at
$ r_{\pm}=M\pm (M^2-Q^2)^{1\over 2}.$
In order to avoid a naked singularity we must have $M\ge Q$. The
Hawking temperature of this hole is  given by
$T = {(r_+-r_-)\over 4\pi {r_+^2}} $ (which goes to zero as
$M\rightarrow Q$ and the entropy is again given by the quarter of
the area of the horizon $S={1\over 4}4\pi (2M)^2=4\pi M^2$ as in
the Schwarzschild case. In the limit $M\rightarrow Q$, the two
horizons become degenerate and the metric of this extremal hole is
$$
ds^2=-(1-{M\over r})^2dt^2+{dr^2\over (1-{M\over r})^2}+r^2 d\O_2.
\eqn\extreme
$$
Although the limiting temperature of the RN black hole in the
extremal limit is zero, purely geometrical considerations of
the extremal hole metric itself indicate that the temperature of
this extremal hole is arbitrary and that its entropy is zero [\hor]
even though the area of the horizon is non-zero. This seems to be
rather puzzling from the thermodynamic point of view.
\foot{We wish to thank L. Susskind for pointing this out.}
We shall see below that the calculation of the contribution of the
scalar fields to the entropy sheds some light on this issue.

For the non-degenerate case, the optical volume is
$$
\eqalign{V^{rn}_3=&4\pi\int_{r_++\e}^R {r^6dr\over
(r-r_+)^2(r-r_-)^2}
=4\pi [{R^3\over 3} +2MR^2+(3r_+^2+4r_+r_-+3r_-^2)R \cr
&+{r_+^6\over (r_+-r_-)^2\e} +{r_-^6\over (r_+-r_-)^3}
+{2r_+^5 (2r_+-3r_-)\over (r_+-r_-)^3}\ln{R-r_+ \over \e}\cr
&+{2r_-^5(3r_+-2r_-)\over (r_+-r_-)^3}\ln {R-r_-\over r_+-r_-}
-r_+({13\over 3}r_+^2+5r_+r_-+3r_-^2)\cr
&+O(R^{-1})+O(\e)].}
\eqn\rnvol
$$
Substituting this in \fourdfree, we have the expressions for the
(quantum corrections to the) free energy and hence also the entropy
in this space. The leading divergence is again linear (or quadratic
in the proper cutoff) and there is an additional logarithmic
divergence. However we also see the appearance of inverse powers of
the difference in the two horizon radii. This clearly implies that
the extremal limit is very singular. Indeed this is confirmed by a
direct calculation of the extremal black hole free energy and
entropy.

{}From \extreme\  we have for the optical volume
$$
\eqalign{V^{ext}_3=&\int_{M+\e}^R {r^6dr\over (r-M)^4}
=4\pi [{R^3\over 3}+2MR^2+10M^2R+{M^6\over 3\e^3}+{3M^5\over\e^2}\cr&
+{15M^4\over\e}+20M^3\ln{R-M\over\e}-{37\over 3}M^3
+O(R^{-1})+O(\e)].}
\eqn\extvol
$$
Here we see the appearance of cubic and quadratic divergences.
Clearly the thermodynamics of the extremal limit is not well-defined
since although the linear and logarithmic divergences
may be absorbed into the renormalization of $G_N$ [\suss] and the
coefficients of higher powers of curvature in the expansion of
the effective action, this will not be the case
for these higher order divergences. {\it The point is that in the
limit $M\rightarrow Q$ the temperature $ T_H\simeq (r_+-r_-)
\rightarrow 0$ so that the free energy (see \fourdfree\ ) goes to
zero while the entropy correction is logarithmically divergent.
However in the extremal case \extreme\ the temperature is arbitrary
[\hor] so that both the free
energy and the entropy will diverge cubically.}
This suggests therefore that the thermodynamics of the extremal RN
black hole \extreme\ as opposed to the limiting case of the RN
black hole \rn\ is not well-defined.

Our last example is the dilaton black hole [\gm]\foot{This case has
been
discussed using a different method in [\ghm].}. The metric is
given in this case by
$$
ds^2=- (1-{2M\over r})dt^2 + {dr^2\over (1-{2M\over r})}+r(r-a)d\O_2,
\eqn\dil
$$
where $a$ is a constant. The corresponding optical volume is
$$
\eqalign{V_3&=4\pi\int^R_{2M+\e}{r(r-a)\over (1-{2M\over r})^2}dr
\cr
&=4\pi[{R^3\over 3 }+(2M-{a\over 2})R^2
+4M(3M-a)R+{8M^3(2M-a)\over\e}\cr
&+4M^2(8M-3a)\ln{R-2M\over\e}-M^2({104M\over
3}-10a)+O(R^{-1})+O(\e)].}
\eqn\dila
$$
As in the Schwarzschild case, here too there are linear as well as
logarithmic divergences and again one may argue following [\suss]
that the former can be absorbed in a renormalization of $G_N$.
In the extremal limit ($M\rightarrow {a\over 2}$), the ``classical"
entropy ($S_{cl}={A\over 4}=2\pi M(2M-a)$ [\preskill]) vanishes and
so does the
linear divergence. However the logarithmic divergence remains.

{}Finally let us point out that our thermodynamic entropy calculation
has a
bulk contribution  in all finite mass black hole cases.
Thus  (unlike in Rindler space [\suss,\call]) this cannot
be identified with the microscopic entropy which is expected to be
proportional to the area of the horizon [\bom,\sred].

\ack
SdeA would like to acknowledge the award of a Japan Society for the
Promotion of Science  fellowship and the hospitality of Profs.
K. Higashijima and E. Takasugi at Osaka University where this
investigation was begun. One of us (SdeA) would also like to thank
Doug MacIntire for discussions. This work is partially supported by
the
Department of Energy contract No. DE-FG02-91-ER-40672.

In the course of writing this paper, we received a preprint
(A. Ghosh and P. Mitra, preprint SISSA-172/94/EP, hep-th/9411128)
in which the entropy for the Reissner-Nordstr\"om black hole was
computed.

\refout

\end